\documentclass[prb,twocolumn,showpacs]{revtex4} 
\usepackage{graphicx}
\usepackage{amssymb,amsmath}
\begin{document}

\title{ A consistent statistical treatment of the renormalized mean - field t-J model}

\author{Jakub J\c{e}drak}
\email{jedrak@th.if.uj.edu.pl}
\author{Jozef Spa\l ek}
\email{ufspalek@if.uj.edu.pl}
\affiliation{Marian Smoluchowski Institute of Physics, Jagiellonian University, Reymonta 4, PL-30059 Krak\'ow, Poland}

\date{\today}

\begin{abstract}
A variational treatment of the Gutzwiller - renormalized t-J Hamiltonian combined with the mean-field (MF) approximation is proposed, with a simultaneous inclusion of additional consistency conditions. Those conditions guarantee that the averages calculated variationally coincide with those calculated from the self-consistent equations.  This is not ensured \textit{a priori} because the effective Hamiltonian contains renormalization factors which depend explicitly on the mean-field averages. 
A comparison with previous mean-field treatments  is made for both  superconducting  (d-RVB) and  normal states and encompasses calculations of both the superconducting gap and the renormalized hopping amplitudes, as well as the electronic structure. The $C_{4v}$-symmetry breaking in the normal phase - the Pomeranchuk instability (PI) - is also analyzed.

\end{abstract}

\pacs{71.27.+a, 74.72.-h, 71.10.Fd}
\maketitle



t-J model \cite{Spalek Oles} is regarded to reflect some of the essential physics of strongly correlated copper $3d_{x^2 - y^2}$ states in high-temperature superconductors.\cite{Lee} 
In this model, the correlated hopping of electrons reduces strongly their band energy, so the latter, for the doping $ x \leq 0.1$, becomes comparable to the real-space pairing part induced by the kinetic exchange.\cite{Spalek didactical tJ}
However, the analytical solutions of t-J model are limited to the special cases for the one-dimensional system.\cite{Sarben Sarkar}
Under these circumstances, we have to resort to either exact diagonalization,\cite{Freericks Falicov} which is limited to small cluster systems  or to the approximate methods. The latter include renormalization group,\cite{Falicov Berker}  variational approach based on the Gutzwiller - projected wavefunctions (either treated within Monte Carlo 
techniques or by Gutzwiller approximation\cite{Edegger}) and  various versions of the  slave - boson approach.\cite{Lee} 
Each of these methods seizes some of the principal features of these quasi-two-dimensional correlated states, although no  coherent picture has emerged as yet. \\

 In this paper we concentrate on the Gutzwiller renormalized mean-field (MF) theory for the t-J Hamiltonian  and  formulate a variational procedure, with the additional conditions ensuring the self-consistency of the whole approach.  Implementing such procedure is essential (if not indispensable) for obtaining reliable results of the MF type.  
  It is reassuring that some of the quantities such as the RVB gap magnitude or the hopping correlations (bond-parameter) do not change appreciably with respect to the earlier results,\cite{Marcin Polonica} 
whereas the others, such as the single-particle electronic structure, are altered remarkably. Furthermore, we illustrate  the basic nontriviality of our approach on the example of the so-called  Pomeranchuk instability discussed recently.\cite{H. Yamase and H. Kohno} 

We start with the t-J model in its simplest form,\cite{Spalek Oles, Lee}
\begin{equation}
 \hat{H}_{tJ} = \hat{P} \big(\sum_{i,j,\sigma} t_{ij} c_{i \sigma}^{\dag} c_{j \sigma} +  \sum_{\langle i j \rangle} J_{ij}~  \mathbf{S}_{i}\cdot \mathbf{S}_{j} -\mu \sum_{i,  \sigma} \hat{n}_{i \sigma}\big) \hat{P},
\label{t-J exact}
\end{equation}
where $\hat{P}$ labels the Gutzwiller projector which guarantees that no doubly occupied sites are present. The projected operators and the model parameters have the standard meaning.\cite{Spalek Oles}

 To proceed further, effective mean-field renormalized  Hamiltonian is introduced  \cite{Edegger, Marcin Polonica, ZGRS, The ladder of Sigrist, Didier, Marcin 1, LiZhouWang, Ogata Himeda} which is taken in the following form 
\begin{eqnarray}
\hat{H} &=&  \sum_{\langle i j \rangle \sigma}\big (t_{ij} g^{t}_{ij} c_{i \sigma}^{\dag} c_{j \sigma} +  \text{H.c.} \big) -\mu  \sum_{i \sigma} c^{\dag}_{i \sigma} c_{i \sigma}  \nonumber \\ &-&  \sum_{\langle i j \rangle \sigma} \frac{3}{4} J_{ij} g^{J}_{ij} (\chi_{ji}  c_{i \sigma}^{\dag} c_{j \sigma} + \text{H.c.} - |\chi_{ij}|^{2}) \nonumber \\ &-& \sum_{\langle i j \rangle \sigma} \frac{3}{4} J_{ij} g^{J}_{ij} (\Delta_{ij}   c^{\dag}_{j \sigma}c_{i -\sigma}^{\dag} + \text{H.c.}   - |\Delta_{ij}|^{2}).
\label{ren tJ Ham}
\end{eqnarray}
In the above expression,  $c_{i \sigma}^{\dag} $ ($ c_{j \sigma}$) are ordinary fermion creation (annihilation) operators, $\chi_{ij} = \langle  c_{i \sigma}^{\dag}  c_{j \sigma} \rangle$, and $\Delta_{ij} = \langle  c_{i -\sigma}  c_{j \sigma} \rangle = \langle  c_{j -\sigma}  c_{i \sigma} \rangle$ are respectively, the hopping amplitude (bond-parameter) and the RVB gap parameter, both taken for nearest neighbors $\langle i j \rangle $. 
The \textit{renormalization factors} $g^{t}_{ij} $ and $g^{J}_{ij} $ result from the Gutzwiller ansatz. The exchange part  ($ \mathbf{S}_{i}\cdot \mathbf{S}_{j} $)  has been decoupled in the Hartree-Fock-type approximation and incorporates as nonzero all above bilinear averages  obtained according to the prescription 

\begin{equation}
 \hat{O}_{\kappa}\hat{O}_{\gamma} \to     \hat{A}_{s} A_{t}  + A_{s} \hat{A}_{t } - A_{s } A_{t }, 
\label{AB BF}
\end{equation} 
where  $t = t(\kappa, \gamma)$ etc. and for any operator $\hat{A}$
\begin{equation}
A = \langle \hat{A} \rangle  \equiv  \text{Tr}[\hat{A}  \hat{\rho}], 
\label{general BdG s-c}
\end{equation}
with $ \hat{\rho}$ being the density matrix for the mean-field Hamiltonian to be determined.
By taking the step from (\ref{t-J exact}) to (\ref{ren tJ Ham}) we introduce essentially a non-Hartree-Fock-type of approximation, which differs from (\ref{AB BF}) due  to the presence of  $g^{t}_{ij} $ and $g^{J}_{ij} $ factors. 
Therefore, we may not be able use e.g. the density operator of the form  $ \hat{\rho} = Z^{-1}e^{-\beta \hat{H} }$, $Z=\text{Tr}[e^{-\beta \hat{H} }]$, as a proper grand-canonical trial state in the frame of variational principle based on the  Bogoliubov  inequality,\cite{Feynman} since then the self-consistency of the approach (expressed by Eq.(\ref{general BdG s-c})) may be violated. This is the reason, why in most of the previous mean field treatments, e.g. \cite{Marcin Polonica, The ladder of Sigrist, Didier, Marcin 1},  the standard procedure encompasses diagonalizing of  the bilinear Hamiltonian (\ref{ren tJ Ham}), and subsequently solving of the self-consistent (\textit{s-c})  Bogoliubov-de Gennes (BdG) equations for $\chi_{ij}$, $  \Delta_{ij}$ and $\mu$. In effect, this procedure does not refer to any variational scheme. 

The solution based solely on the \textit{s-c}  BdG equations, although  acceptable, may not be fully satisfactory.
This is because in the present  situation we build up the entire description  on the basis of MF Hamiltonian and hence we should  proceed in a direct analogy to the exact (non-MF) case. Namely, our approach is based on the maximum entropy principle.\cite{Jaynes} Such starting point  provides us with a general variational principle, which may differ from that of Bogoliubov and Feynman.\cite{Feynman} In other words, the  value of the appropriate  functional is minimized, with the self-consistency of the whole approach being preserved at the same time.\cite{JJJS_arx_0} 

To tackle the situation, we define an effective Hamiltonian $\hat{H}_{\lambda}$ containing additional  constraints, that is of the form
\begin{eqnarray} 
\hat{H}_{\lambda} = \hat{H} &-& \sum_{i} \lambda^{(n)}_{i}\big( \sum_{\sigma} c_{i \sigma}^{\dag}  c_{i \sigma} - n_{i}\big) \nonumber \\ &- &  \sum_{\langle i j \rangle \sigma}  \big( \lambda^{\chi}_{ij}  (c_{i \sigma}^{\dag}  c_{j \sigma} - \chi_{ij}) + \text{H.c.} \big) \nonumber \\ &- &  \sum_{\langle i j \rangle \sigma} \big( \lambda^{\Delta}_{ij}  (c_{i - \sigma}  c_{j \sigma} - \Delta_{ij})  + \text{H.c.} \big),
\label{MF tJ lambda term}
\end{eqnarray}
where the Lagrange multipliers  $ \lambda^{(n)}_{i}$, $\lambda^{\chi}_{ij} $, and $\lambda^{\Delta}_{ij} $  play the role of molecular fields. Moreover, the parameters $\chi_{ij} $,  $\Delta_{ij} $, and $n_i = \sum_{\sigma} \langle  c_{i \sigma}^{\dag}  c_{i \sigma} \rangle $ coincide with those which appear in the renormalization factors  $g^{t}_{ij}$ and $g^{J}_{ij}$, and which are taken in the form  \cite{Marcin Polonica, Marcin 1}  
\begin{equation}
 g^{t}_{ij} = \sqrt{\frac{4x_i x_j(1-x_i)(1-x_j)}{(1-x_i^{2})(1-x_j^{2}) + 8(1-x_i x_j)|\chi_{ij}|^{2} + 16|\chi_{ij}|^{4}}}, \nonumber
\label{gt ren fac x chi delta}
\end{equation}
\begin{equation}
 g^{J}_{ij} = \frac{4(1-x_i)(1-x_j)  }{(1-x_i^{2})(1-x_j^{2}) + 8x_i x_j\beta^{-}_{ij}(2) + 16\beta^{+}_{ij}(4)},
\label{gJ ren fac x chi delta}
\end{equation}
with  $x_i \equiv 1 - n_i $, $\beta^{\pm}_{ij}(n)= |\Delta_{ij}|^{n}\pm|\chi_{ij}|^{n}$. 

When solving the model on a square lattice and in the spatially homogeneous case, there appear thus five mean fields,  
$ \vec{A} \equiv (n, \chi_x, \chi_y, \Delta_x, \Delta_y)$, with $\chi_{\tau} = \chi_{ij}, \Delta_{\tau} = \sqrt{2} \Delta_{ij}$, ($ \langle ij \rangle || \tau$, $\tau = x, y$;  as well as the same number of the corresponding Lagrange multipliers, $\vec{\lambda} \equiv (\lambda, \lambda^{\chi}_x, \lambda^{\chi}_y, \lambda^{\Delta}_x, \lambda^{\Delta}_y)$, where  $\lambda^{\chi}_{\tau} =\lambda^{\chi}_{ij}, \lambda^{\Delta}_{\tau} = \sqrt{2} \lambda^{\Delta}_{ij}$. Both $\vec{A}$ and $\vec{\lambda}$ are assumed to be real. Apart from that, for given $n$ we have to determine the chemical potential $\mu$. The first step is the diagonalization of $\hat{H}_{\lambda}$ via Bogoliubov-Valatin transformation, which yields  
\begin{equation} 
\hat{H}_{\lambda} =  \sum_{\mathbf{k}} E_{\mathbf{k}} (\hat{\gamma}^{\dag}_{\mathbf{k}0}\hat{\gamma}_{\mathbf{k}0} + \hat{\gamma}^{\dag}_{\mathbf{k}1}\hat{\gamma}_{\mathbf{k}1}) + \sum_{\mathbf{k}} (\xi_{\mathbf{k}} - E_{\mathbf{k}}) + C,
\label{MF tJ BCS like}
\end{equation}
with $E_{\mathbf{k}} = \sqrt{\xi^2_{\mathbf{k}} + D^2_{\mathbf{k}}}$, $D_{\mathbf{k}} = \sqrt{2}\sum_{\tau}D_{\tau}\cos(k_{\tau})$, and $\xi_{\mathbf{k}} = -2\sum_{\tau}T_{\tau}\cos(k_{\tau})  -\mu - \lambda$. Also,
\begin{equation} 
T_{\tau} = -t_{1\tau} g^{t}_{1\tau} + \frac{3}{4}J_{\tau} g^{J}_{\tau}\chi_{\tau} + \lambda^{\chi}_{\tau}, ~~~~D_{\tau} =  \frac{3}{4}J_{\tau} g^{J}_{\tau}\Delta_{\tau} + \lambda^{\Delta}_{\tau}, 
\label{T tau}
\end{equation} 
\begin{equation} 
\frac{C}{ \Lambda } =\lambda n + \sum_{\tau} \big( \frac{3}{4} J_{\tau} g^{J}_{\tau} (2 \chi^{2}_{\tau} + \Delta^{2}_{\tau}) + 4 \chi_{\tau} \lambda^{\chi}_{\tau} + 2\Delta_{\tau}\lambda^{\Delta}_{\tau}\big).
\label{C equation}
\end{equation}
For the sake of simplicity, we have included only the hopping between the nearest neighbors, although the generalization to the case with more distant hopping does not pose any principal difficulty. We define next the generalized Landau functional, $\mathcal{F} \equiv -\beta^{-1} \ln( \text{Tr} [e^{-\beta \hat{H}_{\lambda}  }])$, which here takes the form
\begin{equation}
\mathcal{F}(\vec{A}, \vec{\lambda}) =  C + \sum_{\mathbf{k}}\big(  (\xi_{\mathbf{k}} - E_{\mathbf{k}}) - \frac{2}{\beta} \ln\big(1 + e^{-\beta E_{\mathbf{k}}}\big)\big),
\label{mathcalF tJ BCS like}
\end{equation}
with inverse temperature $\beta = 1/k_{B}T$. 
The equilibrium values of $\vec{A}= \vec{A}_0$, $\vec{\lambda} = \vec{\lambda}_0$ are  the solution of the set of  equations
\begin{equation}
\nabla_{A} \mathcal{F} =0, ~~ ~~ \nabla_{\lambda} \mathcal{F} =0, 
\label{derivative of mathcalF A, lambda}  
\end{equation}
for which (\ref{mathcalF tJ BCS like}) reaches its minimum. This step is equivalent to the maximization of the entropy with the constraints.\cite{JJJS_arx_0}
Also, the grand potential $\Omega$ and the free energy $F$ are defined respectively as 
$\Omega(T, V, \mu) = \mathcal{F}(T, V, \mu; \vec{A}_0(T, V, \mu),  \vec{\lambda}_0(T, V, \mu)  )$, and $F = \Omega + \mu N$.
Note, that by taking the derivatives  with respect to $\vec{\lambda}$ only, and subsequently putting $\vec{\lambda}= \vec{0}$, the results reduce to the standard BdG self-consistent equations. 

Even though the present method can be regarded as natural within the context of statistical mechanics,  to the best of our knowledge, it has not been utilized, in the form presented here,  in the context of condensed matter physics problems. Also, in this respect, our approach unifies individual features of the self-consistent variational MF treatments developed earlier\cite{Buenemann Gebhard Thul,Ogata Himeda, LiZhouWang,Wang F C Zhang}, which in the $T=0$ limit can be obtained as particular cases. Parenthetically,  the present method, together with the Gutzwiller approximation, provides also a natural justification of some aspects of the  slave-boson saddle-point approach, as some of the constraints coincide in both methods. 

 We solve numerically first the system of equations (\ref{derivative of mathcalF A, lambda}) on the lattice of  $ \Lambda = 128 \times 128 $ sites, using the periodic boundary conditions and taking the parameters $J_x = J_y = J = 1$, $t_x = t_y = -3 J$, and for low temperature $k_B T / J = 0.002$ for the filling  $n = 7/8 = 0.875$. Both the d-wave superconducting resonating valence bond (d-RVB)  and the isotropic normal  (N) solutions are analyzed. The self-consistent variational results (denoted as \textit{var}) obtained here and those obtained from BdG equations  are compared in Tables I and II. One sees that our value of the low-temperature free energy (per site), (c.f. Table I) in the d-RVB phase is slightly better than the previous  estimates,\cite{Marcin Polonica, Marcin 1} albeit not much ($-1.3661$ for \textit{var}, as compared to $ -1.3647$ for \textit{s-c}). It is slightly higher than that of the Variational Monte Carlo, which is $E_{VMC}/J = -1.3671$,  c.f.\cite{Marcin 1} Also the isotropic staggered-flux  (SF) phase has been found unstable against N state within both methods at this filling. In Table II we display microscopic quantities characterizing each solution in that case and compare them with those obtained within standard \textit{s-c} treatment. The differences are more pronounced for the RVB state. 

\begin{center} 
Table I. Comparison of the values of the thermodynamic potentials (per site). $\tilde{\Omega}$ ($F$) stands for  $  \Omega -\lambda N$ ($\Omega + \mu N$)  for  \textit{var} and $\Omega_{s-c}$ ($\Omega_{s-c} +\mu_{s-c} N$) for \textit{s-c} methods, respectively.\\ 
\begin{tabular}{c c c c c}\hline \hline
Therm. Pot. &  \textit{var} RVB  &   \textit{s-c} RVB   & \textit{var} N  & \textit{s-c} N               \\  \hline 
$\Omega/\Lambda $  &  -5.75856  & -  & -6.25862  &  - \\  
$\tilde{\Omega}/\Lambda$   & -1.07648 & -1.03575 &  -1.11823 &  -1.08025\\  
 $F/\Lambda $   & -1.36614  & -1.36471 & -1.2955672 &  -1.2955671\\  
\hline \hline
\end{tabular}
\end{center}
  
\begin{center}
Table II. Values of chemical potentials and MF parameters.
$\tilde{\mu} $ stands for $\lambda + \mu $ (\textit{var}), and for $\mu_{s-c}$ (\textit{s-c}).\\
\begin{tabular}{c c c c c}\hline \hline

Variable &  \textit{var} RVB &  \textit{s-c} RVB    & \textit{var} N  &  \textit{s-c} N  \\  \hline 
$\mu$  &  5.01989  & -  &  5.67206   &  -   \\  
$\lambda $   & -5.35094  & -  &  -5.87473  & -  \\  
$\tilde{\mu}$   & -0.33105  &  -0.37595  & -0.20267    & -0.24608   \\   
$\chi_x = \chi_y$   &  0.18807  & 0.19074  &  0.20097  & 0.20097 \\  
$\lambda^{\chi}_{x} = \lambda^{\chi}_{y}$   & -0.16985   & -  & -0.18369  &  -\\
$\frac{\Delta_x}{\sqrt{2}} = - \frac{\Delta_y}{\sqrt{2}}$   & 0.13199   & 0.12344  & 0.00000  &   0.00000\\ 
 $\frac{\lambda^{\Delta}_{x}}{\sqrt{2}} = - \frac{\lambda^{\Delta}_{y}}{\sqrt{2}}$   & -0.01111   &  - &  0.00000 &  -\\  
\hline \hline
\end{tabular}
\end{center}
\begin{figure}[h]
\begin{center}
\rotatebox{270}{\scalebox{0.32}{\includegraphics{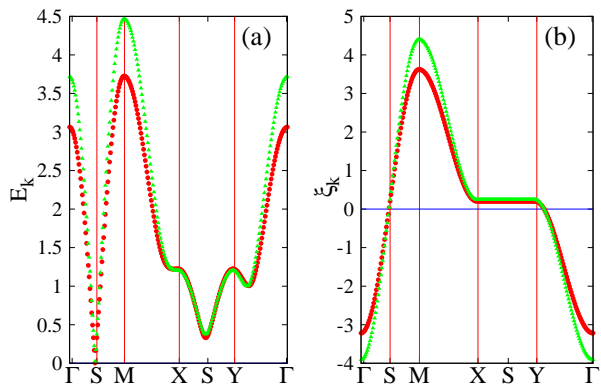}}}
\end{center}
\caption{(Color online) Dispersion relations along the main symmetry lines in the Brillouin zone for a square lattice, of the size $\Lambda_x = \Lambda_y  = 128$, and for the filling $n=0.875$. Left (a):  d-RVB solutions, right (b): N solutions. Triangles - earlier self-consistent  results, circles - the present  method.}
\label{dispertion for 128 RVB}
\end{figure}
For the parameters listed in Tables I and II we have computed the quasiparticle energies in both the d-RVB
and the N states. Those are shown in Fig. 1a-b. The solid circles represent our results, whereas the previous ones \cite{Marcin Polonica} are drawn as triangles.
The energy-dispersion reduction  in our case is connected with  presence of the constraints and results in a decrease of  the bandwidth, which, in turn, is regarded as a sign of enhanced electron correlations. 

\begin{figure}[h!]
\begin{center}
\rotatebox{270}{\scalebox{0.29}{\includegraphics{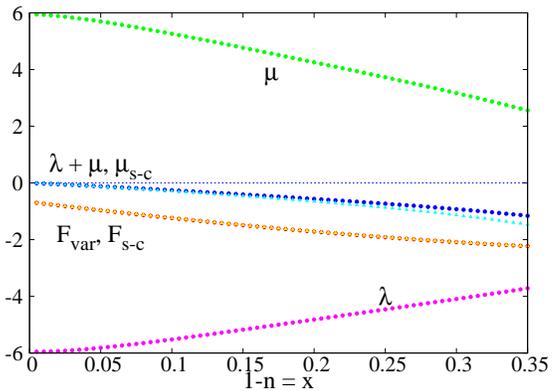}}}
\end{center}
\caption{(Color online) Doping dependences of the free energy ($F_{var}, F_{s-c}$), the chemical potentials   $\mu$, $\mu_{s-c}$ as well as that of $\lambda$ and $\lambda + \mu$ for the d-RVB state, both within the present (\textit{var}) and the standard (\textit{s-c}) methods.} 
\label{doping dependance of F mu etc for the plane}
\end{figure}

\begin{figure}[h!]
\begin{center}
\rotatebox{270}{\scalebox{0.32}{\includegraphics{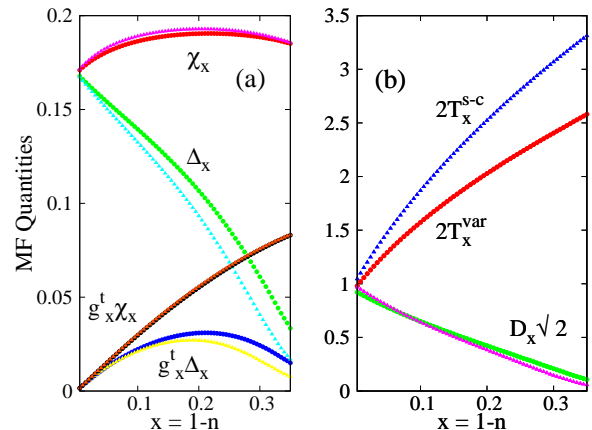}}}
\end{center}
\caption{(Color online) Left: Doping dependence of the bond-order parameters $\chi_x=\chi_y$,  the superconducting order parameters $\Delta_x = - \Delta_y$,  their renormalized counterparts $g^{t}\chi_x=g^{t}\chi_y$ and $g^{t}\Delta_x = - g^{t}\Delta_y$, as well as (right) of the quantities $2 T_x = 2 T_y$ and $\sqrt{2} D_x = -\sqrt{2} D_y$ of Eq. (\ref{T tau}), both for the \textit{s-c} (triangles) and the \textit{var} (circles) methods.}
\label{x dependence of MFs for RVB with T and D}
\end{figure}

After testing the feasibility of our approach for fixed doping $x$, we now discuss systematic changes appearing as the function of $x$, as shown in Figs. \ref{doping dependance of F mu etc for the plane} and \ref{x dependence of MFs for RVB with T and D}.

 We emphasize, the chemical potential $\mu$ is the first derivative of $F/\Lambda$ with respect to $n$ (c.f. Fig.\ref{doping dependance of F mu etc for the plane}), unlike in  some of the previous mean-fields treatments \cite{Marcin Polonica, The ladder of Sigrist, Didier, Marcin 1} (c.f.  however Ref. \cite{ZGRS, Buenemann Gebhard Thul}). This is also the reason why we differentiate between $\mu$ and $\tilde{\mu} \equiv \mu + \lambda$, even in the case of the spatially homogeneous solution. 
The doping dependence of  other relevant  MF quantities is shown  in Fig.\ref{x dependence of MFs for RVB with T and D}. The results are again close to those obtained from the BdG procedure, except for $T_{\tau}$, (Fig.\ref{x dependence of MFs for RVB with T and D}(b)), which enter  the quasiparticle energies.

So far we have focused on MF solutions with the symmetry between $x$ and $y$ directions on the square lattice.\cite{Ladder of J} However, a spontaneous breakdown of this  equivalence of the $x$- and $y$- directed correlations is possible already in the normal phase and is called \textit{the Pomeranchuk instability} (PI), \cite{H. Yamase and H. Kohno} that manifests itself by lowering of the discrete $C_{4v}$ symmetry of the Fermi surface. 

\begin{figure}[h!]
\begin{center}
\rotatebox{270}{\scalebox{0.32}{\includegraphics{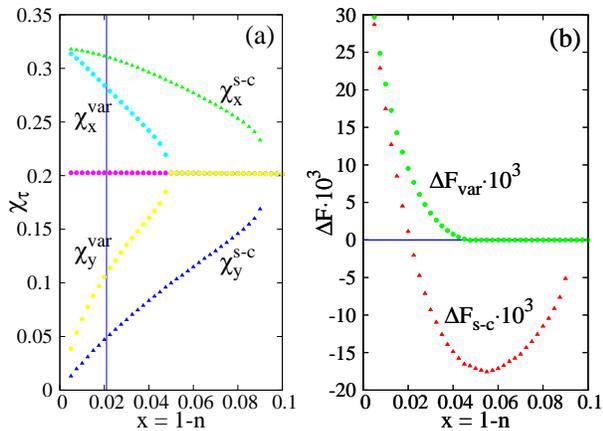}}}
\end{center}
\caption{(Color online) Doping dependence of bond-order parameters $\chi_x$ and $\chi_y$ (left), and the free-energy differences $\Delta F$ (right) both for $x$-$y$ symmetric (N) and the Pomeranchuk (x-y symmetry-broken) states  (PI)  within both the present(\textit{var}, filled circles) and the standard  (\textit{s-c}, triangles)  methods, respectively. The vertical line marks  the phase transition within the \textit{s-c} method. For details, see main text.}
\label{doping dependance of F and chi PI}
\end{figure}

In Fig. \ref{doping dependance of F and chi PI} (a) the doping dependence of the bond-order parameters $\chi_x$ and $\chi_y$ are displayed for  the $x$-$y$ symmetric (N) and the  symmetry-broken (PI) solutions, both within  our ($\chi^{var}_{\tau}$) and the standard ($\chi^{s-c}_{\tau}$) methods. Within the \textit{s-c} scheme, PI solution is found up to $x \approx 0.091$. However, a comparison of the respective free-energy differences, $\Delta F_{s-c}\equiv F_{s-c}^{N} - F_{s-c}^{PI}$  and $\Delta F_{var}\equiv F_{var}^{N} - F_{var}^{PI}$,  (cf. Fig. \ref{doping dependance of F and chi PI} (a)) reveals that this solution becomes unstable against N state for  $x \approx 0.021$, thus the phase transition is certainly discontinuous. On the other hand, within our variational treatment  the  PI solution does not exist for $x > x^{var}_{c}\approx 0.044$, where $\Delta F_{var} \approx 0$, in qualitative agreement with what is expected for the continuous phase transition.
 From this analysis it is clear that the two methods of approach (\textit{s-c}, \textit{var}) yield qualitatively different  predictions for PI.

In summary, we have introduced self-consistency constraints required within the variational mean-field approach to the Gutzwiller-renormalized mean-field t-J model. Such consistency conditions are indispensable from the basic statistical-mechanical point of view. Undertaking such a step  results in consistent evaluations of the  thermodynamic quantities, which in the present method are determined from the generalized Landau functional. A detailed comparison with the standard mean-field solution based on Bogoliubov-de Gennes self-consistent equations (i.e. that without constraints) is provided.  Our method  introduces quantitative and, in some cases, even  qualitative corrections to the standard mean-field results. Other mean-field states such as flux phases or antiferromagnetism can be treated in the same manner.

The authors are very grateful to  Marcin Raczkowski and Andrzej Kapanowski for valuable comments and technical help. All the numerical computations were performed using GSL (Gnu Scientific Library) efficient procedures. 
The work was supported by the Grant No. N N 202 128 736 from the Ministry of Science and Higher Education, as well as by the National Network \textit{Strongly Correlated Electrons} and the European COST Network (ECOM). \\ 


\begin{thebibliography}{00}
\bibitem{Ladder of J}
The modifications introduced in our approach are relevant also  for a two-leg ladder systems,\cite{The ladder of Sigrist}  but the results are not discussed here.
\end{thebibliography}
\end{document}